\documentclass[]{aa} 

\usepackage{graphicx}
\usepackage{txfonts}
\usepackage{natbib}
\usepackage{longtable}

\begin{document}

\title{The effect of temperature mixing on the observable ($T$,
$\beta$)-relation of interstellar dust clouds
}

\author{M.     Juvela\inst{1},
        N.     Ysard\inst{1}
        }

\institute{
Department of Physics, P.O.Box 64, FI-00014, University of Helsinki,
Finland, {\em mika.juvela@helsinki.fi}
}

\authorrunning{M. Juvela et al.}

\date{Received September 15, 1996; accepted March 16, 1997}

\abstract
{
Detailed studies of the shape of dust emission spectra are possible
thanks to the current instruments capable of simultaneous observations
in several sub-millimetre bands (e.g., Herschel and Planck). The
relationship between the observed spectra and the intrinsic dust grain
properties is known to be affected by the noise and the line-of-sight
temperature variations. However, some controversy remains even on the
basic effects resulting from the mixing of temperatures along the
line-of-sight or within the instrument beam.
}
{
Regarding the effect of temperature variations, previous studies have
suggested either a positive or a negative correlation between the
colour temperature $T_C$ and the observed spectral index $\beta_{\rm
Obs}$. Our aim is to show that both cases are possible and to
determine the principal factors leading to either behaviour.
}
{
We start by studying the behaviour of the sum of two or three modified
black bodies at different temperatures. Then, with radiative transfer
models of spherical clouds, we examine the probability distributions
of the dust mass as a function of the physical dust temperature. With
these results as a guideline, we examine the ($T_C$, $\beta_{obs}$)
relations for different sets of clouds.
}
{
Even in the simple case of models consisting of two blackbodies at
temperatures $T_0$ and $T_0+\Delta T_0$, the correlation between $T_C$
and $\beta_{obs}$ can be either positive or negative. If one compares
models where the temperature difference $\Delta T_0$ between the two
blackbodies is varied, the correlation is negative. If the models
differ in their mean temperature $T_0$ rather than in $\Delta T_0$,
the correlation remains positive.
Radiative transfer models show that externally heated clouds have
different mean temperatures but the widths of their temperature
distributions are rather similar. Thus, in observations of samples of
such clouds the correlation between $T_{\rm C}$ and $\beta_{\rm Obs}$
is expected to be positive. The same result applies to clouds
illuminated by external radiation fields of different intensity. For
internally heated clouds a negative correlation is the more likely
alternative.
}
{
Previous studies of the ($T_{\rm C}$,$\beta$) relation have been
correct in that, depending on the cloud sample, both positive and
negative correlations are possible. For externally heated clouds the
effect is opposite to the negative correlation seen in the
observations. 
If the signal-to-noise ratio is high, the observed
negative correlation could be explained by the temperature dependence of
the dust optical properties but that intrinsic dependence could be
even steeper than the observed one.
}
\keywords{
ISM: clouds -- Infrared: ISM -- Radiative transfer -- Submillimeter: ISM
}

\maketitle
%

\section{Introduction}

Thermal dust emission is increasingly important as a tracer of the
dense interstellar clouds and of the star formation process. Earlier
ground-based and satellite observations already showed that the long
wavelength dust opacity is not constant and the variations are
probably tracing grain coagulation and aggregation processes
\citep{Cambresy2001, delBurgo2003, Kramer2003, Lehtinen2007}. The
sub-millimetre balloon borne experiments PRONAOS and Archeops enabled
the study of the dust spectral index for a large number of
interstellar clouds. By a coverage of the sub-millimetre regime, it
was possible to start to separate the effects of the colour
temperature, $T_{\rm C}$, and of the dust emissivity spectral index,
$\beta$. 
The colour temperature $T_{\rm C}$ can be derived from the fit of
a modified black body curve $B(T_{\rm C}) \nu^{\beta_{\rm Obs}}$ to the
multi-wavelength observations.  The results indicated a negative
correlation between the colour temperature and the spectral index with
$\beta_{\rm Obs}$ increasing in the cold regions \citep{Dupac2003,
Desert2008}.
The results were not universally trusted because the variables are
intrinsically negatively correlated so that a small error in
$\beta_{\rm Obs}$ can be compensated by a small error in $T_{\rm C}$
in the opposite direction. 
Therefore, and because of the other effects discussed in this
paper, one must make the separation between the apparent spectral
index $\beta_{\rm Obs}$ and the intrinsic spectral index $\beta$ of
the dust grains.
In particular, \citet{Shetty2009a} and \citet{Shetty2009b} showed that
a negative correlation between $T_{\rm C}$ and $\beta_{\rm Obs}$ could
result from noise. However, the recent Planck and Herschel satellite
results also show a decreasing $\beta_{\rm Obs}(T_{\rm C})$ relation
that, according to the authors, is too strong to be explained by the
noise alone \citep{Anderson2010, Paradis2010, Veneziani2010,
Planck2011a, Planck2011b}. 
The Herschel bands from 100\,$\mu$m to 500\,$\mu$m cover the peak
of the dust emission. By covering the longer wavelengths from
350\,$\mu$m to millimetre waves, Planck is in principle in a better
position to measure the emission spectral index. However, for an
accurate determination of $\beta$ one also must constrain the
temperature. For this reason the Planck data are being combined with
far-infrared observations, for example the 100\,$\mu$m IRAS data
\citep{Planck2011a, Planck2011b}.

The negative correlation of $T_C$ and $beta_{\rm Obs}$ can be
related to laboratory measurements of some interstellar dust analogues
\citep{Mennella1998, Coupeaud2011} and could be explained by models
developed for the emission of amorphous solids \citep{Meny2007,
Paradis2011}.
However, even if one accepts the negative correlation between
$\beta_{\rm Obs}$ and $T_{\rm C}$ as a real property of the observed
radiation, there are still some hurdles before conclusions can be
drawn regarding the intrinsic properties of the dust grains. The
mixing of different dust temperatures along the line-of-sight, or more
generally within the beam, means that the peak of the dust emission
broadens and that the observed $\beta_{\rm Obs}$ values are reduced.
At the same time, the colour temperature will tend to overestimate the
mass averaged physical dust temperature, possibly resulting in a
serious underestimation of the cloud masses \citep{Malinen2011}. In
this case the observed points would be displaced towards the lower
right in the ($T_{\rm C}$, $\beta_{\rm Obs})$ plane. This has been
interpreted as evidence that the negative correlation between the
observed $T_{\rm C}$ and $\beta_{\rm Obs}$ values also could result
from temperature variations \citep{Shetty2009b}. However,
\citet{Malinen2011} carried out radiative transfer calculations for
turbulent model clouds containing gravitationally bound cores. As long
as the cores were externally heated, the result was an apparent {\em
positive} correlation between $T_{\rm C}$ and $\beta_{\rm Obs}$.

The purpose of this paper is to seek an explanation for this apparent
contradiction and to establish what is the expected behaviour for
dense clouds and why. 

We start by describing our methods and the basic assumptions in
Sect.~\ref{sect:methods} The main results are presented in
Sect.~\ref{sect:results}. We start with two and three layer models
and, in this simple setting, examine the basic effects that the
temperature mixing has on the observable $T_{\rm C}$ and $\beta_{\rm
Obs}$ values. In Sect.~\ref{sect:BE} we carry out radiative transfer
modelling of a series of spherical model clouds to determine what kind
of temperature probability distributions are expected for a collection
of dense clouds. In Sect.~\ref{sect:final} we use this information
together with the two layer models to demonstrate the expected
behaviour for quiescent clouds. Our final conclusions are presented in
Sect.~\ref{sect:discussion}.

\section{The methods} \label{sect:methods}

We examine the relations between the intrinsic dust temperature, $T$,
and opacity spectral index, $\beta$, with the corresponding parameters
derived from the analysis of the observed emission. These are the
colour temperature, $T_{\rm C}$, and the observed spectral index,
$\beta_{\rm Obs}$, both deduced from the shape of the intensity
spectrum. The observed parameters are calculated by fitting a modified
black body fit, $B_{\nu}(T_{\rm C}) \nu^{\beta_{\rm Obs}}$, to
observations in the five Herschel bands at 100\,$\mu$m, 160\,$\mu$m,
250\,$\mu$m, 350\,$\mu$m, and 500\,$\mu$m. We use the monochromatic
intensity values and weighted least squares fits where the same 
relative uncertainty is assumed for all the bands.

The simplest case where the observed spectral index $\beta_{\rm Obs}$
can differ from the intrinsic $\beta$ of the grains, is the two layer
model. The source consists of two layers at different temperature both
possibly still having the same intrinsic $\beta$ value. We will also
investigate cases with three layers, i.e., mixtures of dust at three
temperatures. The emission is assumed to be optically thin so that the
observed intensities are a sum of the emission of the individual
layers. This allows us to compare our findings directly with the
results of \citet{Shetty2009a}. The model parameters that can be varied
are the physical dust temperatures and the relative column densities
of the layers. The value of $\beta$ is assumed to be 2.0 but we will
also briefly look at cases with different $\beta$ values in the
layers.

We will carry out radiative transfer modelling of Bonnor-Ebert spheres
\citep{Bonnor1956}. These calculations provide first information on the
temperature distributions that are likely to be found in real clouds.
The density profiles are calculated for almost critically stable
configurations (stability parameter $\xi$=6.5) with a gas kinetic
temperature of 10\,K. The precise values of these parameters are not
crucial here because the dust temperature distributions depend mainly
on the column density \citep[e.g.][]{Fischera2011}. We start radiative
transfer modelling assuming that the clouds are heated by an external
radiation field corresponding to the interstellar radiation field
(ISRF) model of \citet{Black1994}. The dust properties are those of the
Milky Way dust with $R_{\rm V}$=5.5 \citep{WeingartnerDraine2001}. The
dust temperature profiles are derived with Monte Carlo radiative
transfer calculations \citep{Juvela2003, Juvela2005}. Together with the
density profiles, these can be converted to probability distributions
of the column density as a function of dust temperature. The
Bonnor-Ebert models are discussed in Sect.~\ref{sect:BE}.

\section{The results}  \label{sect:results}

\subsection{Two layer models} \label{sect:two}

\begin{figure}
\centering
\includegraphics[width=8cm]{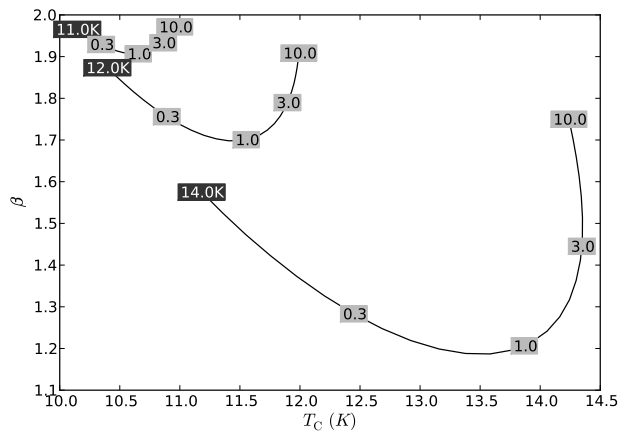}
\caption{
The variation of the observed colour temperature and spectral index in
two layer models (solid curves). The first layer is at fixed
temperature of $T=$\,10\,K. The other layer is at 11\,K, 12\,K, or
14,K (black labels in the figure) and its relative mass is varied
between 0.1 and 10.0 (four values are indicated on the curves). The
intrinsic spectral index of both components is $\beta$=2.0.
}
\label{fig:mixture_1}%
\end{figure}

Figure~\ref{fig:mixture_1} shows the observed temperatures and
spectral indices, $T_{\rm C}$ and $\beta_{\rm Obs}$, for a series of
two layer models. The first dust layer has a temperature of $T$=10\,K
and the other one a temperature of 11\,K, 12\,K, or 14\,K. Both have
an intrinsic opacity spectral index of $\beta$=2.0. In
Fig.~\ref{fig:mixture_1}, the curves correspond to different mass
ratios where the relative mass of the warmer dust is varied between
0.1 and 10.

The correlation between $T_{\rm C}$ and $\beta_{\rm Obs}$ is positive
when the mass ratio of warm to cold dust is close to unity or slightly
higher. When either temperature component is dominating the mass, the
correlation between $T_{\rm C}$ and $\beta_{\rm Obs}$ becomes
negative. If one compares models with the same mass ratios but
different temperature combinations (i.e., 10\,K+11\,K vs. 10\,K+14\,K)
the correlation also is negative.

A two layer model is too simple to be directly compared to actual
observations. Nevertheless, one can appreciate the large range of
possible ($T_{\rm C}$, $\beta_{\rm Obs}$) combinations that are
theoretically possible. In Fig.~\ref{fig:mixture_1} each curve would
end at a value of $\beta_{\rm Obs}$=2.0 both when the mass ratio goes
to zero or infinity. One notable feature is that for a large
temperature difference, i.e., 10\,K mixed with 14\,K, the curve passes
through a region where $T_{\rm C}$ exceeds the physical temperature of
the both components. Such behaviour was already observed by
\citet{Shetty2009a}.

\begin{figure}
\centering
\includegraphics[width=8cm]{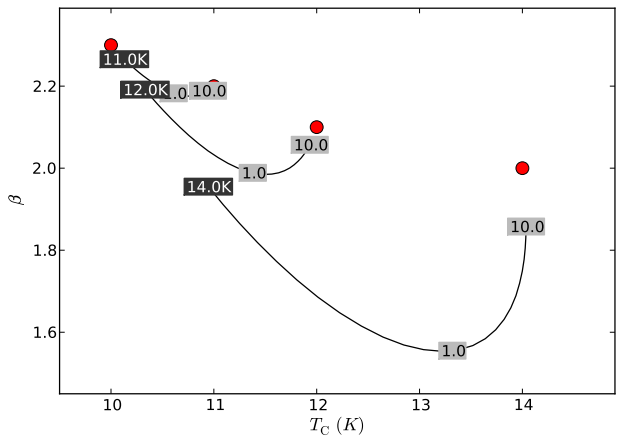}
\caption{
Modification of the two layer model of Fig.~\ref{fig:mixture_1} with
different intrinsic spectral indices. The 10\,K cold component has an
intrinsic spectral index of $\beta=2.3$. The warmer component is at a 
temperature of 11\,K, 12\,K, or 14\,K (as indicated by the black
labels) and has a spectral index $\beta$ of 2.2, 2.1, and 2.0,
respectively. The red circles correspond to these ($T$, $\beta$)
values of the individual dust components.
}
\label{fig:mixture_1x}%
\end{figure}

In Figure~\ref{fig:mixture_1x}, we use different intrinsic $\beta$
values for the two layers with $\beta=$2.3, 2.2, 2.1, and 2.0, for the
temperatures of 10\,K, 11\,K, 12\,K, and 14\,K, respectively. This 
approximates the negative correlation reported in observational
studies \citep[e.g.][]{Dupac2003, Desert2008, Planck2011a}. Compared
to Fig.~\ref{fig:mixture_1}, the curves are tilted to show stronger
negative correlation between $T_{\rm C}$ and $\beta_{\rm Obs}$. 
The difference between $\beta_{\rm Obs}$ and the average intrinsic
$\beta$ is smaller than in Fig.~\ref{fig:mixture_1}. To some extent
this is expected because the warm component, with the smaller $\beta$,
dominates the observed intensities. However, in
Fig.~\ref{fig:mixture_1x} the combination of 10\,K+14\,K has a minimum
at $\beta_{\rm Obs}\sim$1.55, less than 0.8 units below even the
$higher$ of the two $\beta$ values. In Fig.~\ref{fig:mixture_1} the
discrepancy between $\beta_{\rm Obs}$ and the $common$ $\beta$ is
larger.

\subsection{Three layer models}  \label{sect:three}

\begin{figure}
\centering
\includegraphics[width=8cm]{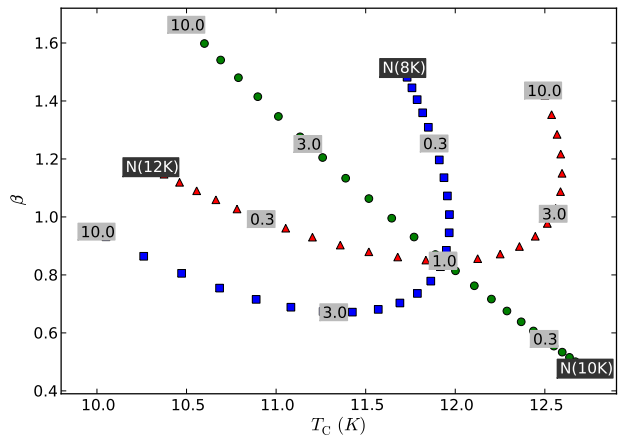}
\caption{
As Fig.~\ref{fig:mixture_1} but for three layers at 8\,K, 10\,K and
12\,K. The central point corresponds to identical mass in each
temperature component and the curves are obtained by scaling the
relative mass of one component at a time. For the component that is
being modified, the temperature is indicated in the figure (black
labels) and the relative mass is shown for a few positions along the
resulting curves (gray labels). All components have the same spectral
index, $\beta=$2.0.
}
\label{fig:mixture_2}%
\end{figure}

We examine models with three temperature layers to make sure that the
qualitative behaviour does not change as one takes the first step
towards more continuous temperature distributions.
Figure~\ref{fig:mixture_2} shows the results for a mixture of dust at
8\,K, 10\,K, and 12\,K. The starting point is a situation where the
three temperature components have equal mass. The relative masses are
then scaled between 0.1 and 1.0, one layer at a time. The first
observation is that the $\beta_{\rm Obs}$ values are now significantly
lower because of the addition of the colder component at 8\,K. 

When the column density of the 8\,K or of the 12\,K component is
varied, the behaviour is qualitatively similar to
Fig.~\ref{fig:mixture_1}. When the mass of the 10\,K component is
varied, the ($T_{\rm C}$, $\beta_{\rm Obs}$) values fall on an almost
straight line. 
The column densities were scaled in multiplicative steps of 1.2 so
that the distance between the plotted symbols corresponds to a
constant step in $log\,N$. The corresponding displacements in the
($T_{\rm C}$, $\beta_{\rm Obs}$) plane are quite regular, deviations
becoming noticeable mainly when the relative mass of the 8\,K layer is
modified. 

If one continues the curves for column density multipliers larger than
10, they will eventually reach the point with $\beta_{\rm Obs}$=2.0
and $T_{\rm C}$ equal to the $T$ of the dominant layer. The
correlation between the $T_{\rm C}$ and $\beta_{\rm Obs}$ values can
again be either positive or negative, depending on which particular
models are being compared. Thus the key question becomes, how the real
clouds differ from each other regarding the range of temperatures and
the relative masses of the temperature components. To answer this
question, we need models where the temperature distributions are solved
self-consistently considering the balance of dust heating and cooling.

\subsection{Bonnor-Ebert spheres} \label{sect:BE}

\begin{figure}
\centering
\includegraphics[width=8cm]{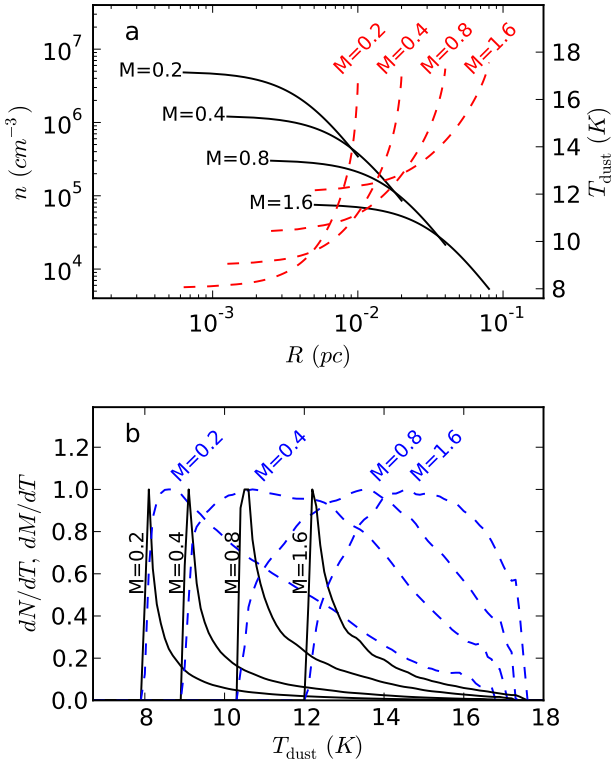}
\caption{
{\em Frame a:} The radial density profiles (solid curves and the scale
on the left) and the temperature profiles (dashed curves and the scale
on the right) for Bonnor-Ebert spheres of 0.2, 0.4, 0.8, and
1.6\,M$_{\sun}$ (curves from left to right).
{\em Frame b:} The probability distributions of the column
density as the function of the dust temperature for a line-of-sight
through the centre of the cloud (solid lines). The curves correspond
to the 0.2, 0.4, 0.8, and 1.6\,M$_{\sun}$ models, from left to right.
The dashed lines show the corresponding probability distributions for
the total dust mass of the model cloud. All the probability
distributions have been normalised to a maximum value of 1.0.
}
\label{fig:Ndist}%
\end{figure}

We solve dust temperatures for a series of Bonnor-Ebert spheres with
masses 0.2, 0.4, 0.8, and 1.6\,$M_{\sun}$ (see
Sect.~\ref{sect:methods}). The first frame of Fig.~\ref{fig:Ndist}
shows the density and temperature profiles of the model clouds. The
temperature at the surface of the clouds is 17--18\,K and the central
temperature decreases from $\sim$12\,K in the 1.6\,$M_{\sun}$ cloud to
$\sim$8\,K in the 0.2\,$M_{\sun}$ cloud. The second frame shows the
probability distributions of the column density on a line-of-sight
through the cloud centre as a function of the temperature.  The shape
of the distributions is rather similar irrespective of the cloud mass
and the probability distributions are only shifted towards lower $T$
as the column density of the model increases, i.e., when the mass of
the Bonnor-Ebert sphere is decreased. The situation is less clear when
we consider the total dust mass. The main effect is still a shift in
the temperature although the skewness of the distribution is larger
for the clouds with the smallest mass. This suggests that the main
variation for these externally heated clouds is in the mean
temperature, not in the magnitude of the temperature variation nor in
the relative mass of the (in relative terms) warm and cold components.

\subsection{The source of a positive correlation} \label{sect:final}

\begin{figure}
\centering
\includegraphics[width=8cm]{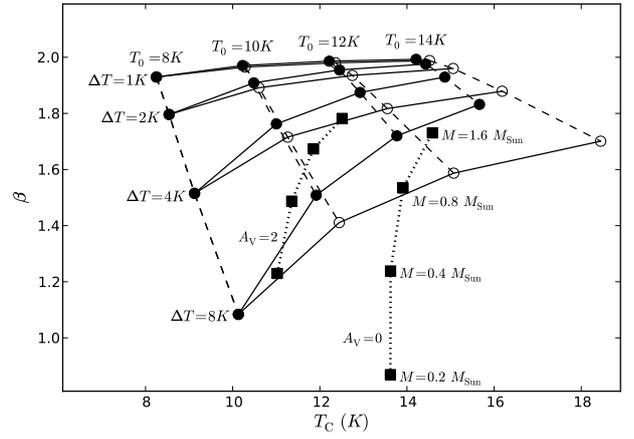}
\caption{
The dependence between $T_{\rm C}$ and $\beta$ for a model of two
layers at temperatures $T_0$ and $T_0+\Delta T$, with a ratio 4:1
between the column densities. Each solid curve corresponds to a single
value of $\Delta T$, with $T_0$=8, 10, 12, and 14\,K along the dashed
lines. The open symbols show corresponding results for column density
ratios 4:1, 3:1, 2:1, and 1:1 for the $T_{\rm 0}$ values of 8, 10, 12,
and 14\,K, respectively.
The two dotted lines and the square symbols show the values for
Bonnor-Ebert spheres of 0.2, 0.4, 0.8, and 1.60\,$M_{\sun}$) with no
external shielding or spheres shielding by $A_{\rm V}=2^{\rm m}$. In
this figure, the 100\,$\mu$m intensities were not used.
}
\label{fig:final}%
\end{figure}

With the knowledge of the temperature distributions found in the
radiative transfer models we can now return to the two layer models
and use them to demonstrate the behaviour of an ensemble of clouds.
The models consist of two layers at temperatures $T_{\rm 0}$ and
$T_{\rm 0}+\Delta T$. 
Because we want to compare these results to the spectra calculated for
the Bonnor-Ebert models and want to avoid the complication of the
contribution of the stochastically heated dust grains to the
100\,$\mu$m intensity, these fits were done without the 100\,$\mu$m
points.
Figure~\ref{fig:final} shows the results for different values of
$\Delta T$, for a set of models with $T_{\rm 0}$=8.0, 10.0, 12.0, or
14.0\,K. A ratio of 4:1 is assumed for the column densities of the
cold and warm components. The same figure also shows the results for
the Bonnor-Ebert spheres. In the two cases, the clouds are illuminated
either by the full ISRF or by the ISRF attenuated by an external layer
of dust with $A_{\rm V}=2^{\rm m}$.

When $\Delta T$ is kept constant but $T_{\rm 0}$ is varied, the
behaviours are qualitatively similar to the Bonnor-Ebert spheres in
that the correlation between $T_{\rm C}$ and $\beta_{\rm Obs}$ is
positive. On the other hand, if $\Delta T$ is varied keeping $T_{\rm
0}$ constant, the correlation becomes negative. The calculations were
repeated with column density ratios 4:1, 3:1, 2:1, and 1:1 for the
$T_{\rm 0}$ values of 8, 10, 12, and 14\,K, respectively (see
Fig.~\ref{fig:final}, the open symbols). This does not have a major
impact on the correlations between the $T_{\rm C}$ and the $\beta_{\rm
Obs}$ parameters.

This key conclusion is that the sign of the correlation entirely
depends on the models one chooses to compare. Therefore, it is
incorrect to interpret the \citet{Shetty2009b} results as a general
proof that the temperature variations always cause a negative
correlation between $T_{\rm C}$ and $\beta_{\rm Obs}$.

\section{Discussion} \label{sect:discussion}

In their article \citet{Shetty2009a} examined the effect of noise and
of line-of-sight temperature variations on the observed $T$ and
$\beta$ values. They concluded that both these factors can produce a
negative correlation between $\beta_{\rm Obs}$ and $T_{\rm C}$ or at
least significantly affect the observed $\beta_{\rm Obs}T_{\rm C}$
relations.
Contrary results were obtained by \citet{Malinen2011} who analysed
synthetic sub-millimetre observations. The cloud model was based on
magnetohydrodynamic simulations where the self-gravity had produced a
number of cores, some with very high column densities. Radiative
transfer modelling was carried out and the synthetic observations were
analysed. The $\beta_{\rm Obs}$ values were always below the intrinsic
$\beta$ but a clear {\em positive} correlation was seen at low
temperatures, i.e., associated with the set of dense cores. The
proposed explanation was that one is again mixing cold and warm dust.
In \citet{Malinen2011} the correlations were derived for all
pixels in the synthetic maps. We present in
Appendix~\ref{sect:collins} further analysis of those data,
concentrating on the locations of the dense cores.

An explanation based on the temperature mixing is, however, 
incomplete because we have seen that the result depends on the way how
the warm and the cold components are mixed. Most importantly, one
needs to know what is the basic difference of the clouds one is
observing. The radiative transfer calculations carried out with
spherical model clouds (Sect.~\ref{sect:BE}) showed that, in the first
approximation, the main difference is their mean temperature. The
shape of the probability distribution for the column density as a
function of temperature remained rather similar and the distribution
was only shifted along the temperature axis (Fig.~\ref{fig:Ndist}b).
In this scenario, the simplest description of an ensemble of clouds is
a set of two layer models with temperatures $T_{\rm 0}$ and $T_{\rm
0}+\Delta T$ where the clouds differ in $T_{\rm 0}$ but not in $\Delta
T$. For unresolved clouds the changes in the shape of the $N(T)$
distribution were more noticeable but still not sufficient to alter
the conclusion. The expected behaviour is a positive correlation
between $T_{\rm C}$ and $\beta_{\rm Obs}$ (Fig.~\ref{fig:final}).

\begin{figure}
\centering
\includegraphics[width=8cm]{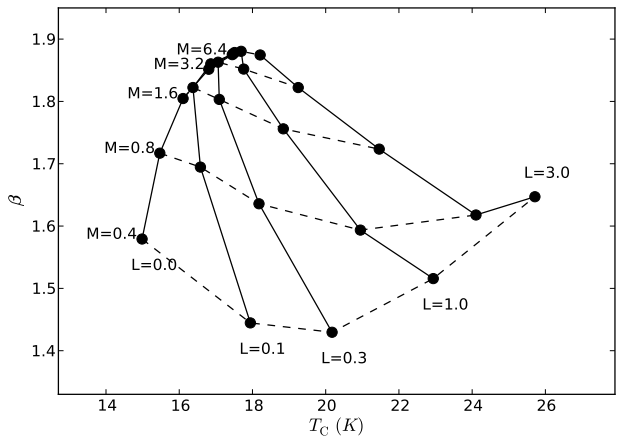}
\caption{
The parameters $T_{\rm C}$ and $\beta$ for a set of internally heated
Bonnor-Ebert models with masses of 0.4, 0.8, 1.6, 3.2, and
6.4\,$M_{\sun}$.  The solid lines connect models with an internal
heating source of given luminosity, 0.0, 0.1, 0.3, 1.0, or 3.0 solar
luminosities. The dashed lines connect models of the same mass. The
parameters have been derived assuming the source in unresolved.
}
\label{fig:internal}%
\end{figure}

If the parameter $\Delta T$ is varied instead, the result is a strong
negative correlation between the temperature and the spectral index. 
\citet{Shetty2009a} did not directly show the $T_{\rm C}(\beta_{\rm
Obs}$) relations as a function of model parameters. Comparing their
Figs. 3 and 4 one can find either positive or negative correlation.
For example, the correlation is negative between the 10\,K+15\,K and
the 10\,K+20\,K cases, precisely when the parameter $\Delta T$ is
being changed.
In \citet{Malinen2011}, the correlation between $T_{\rm C}$ and
$\beta_{\rm Obs}$ also became predominantly negative when internal
heating was applied to the sources. This is schematically consistent
with the above picture. When cores are heated by sources of different
luminosity, the temperature gradients will be different. In other
words, one is primarily modifying the $\Delta T$ parameter. In
Fig.~\ref{fig:final}, this would correspond to the cores being
distributed in a direction perpendicular to the lines of constant
$\Delta T$ \citep[see also Fig. 8 of][]{Shetty2009a}.
To illustrate this further, Fig.~\ref{fig:internal} shows results for
a set of Bonnor-Ebert models with internal heating. The heating source
is a 5800\,K black body that is placed at the centre of the
Bonnor-Ebert sphere. The figure shows the results for the cores
assuming they are not resolved in the observations. As the luminosity
of the point source increases, the correlation between $T_{\rm C}$ and
$\beta_{\rm Obs}$ becomes negative. This is true whether one is
comparing models of different mass or models with different source
luminosity.

\begin{figure}
\centering
\includegraphics[width=8cm]{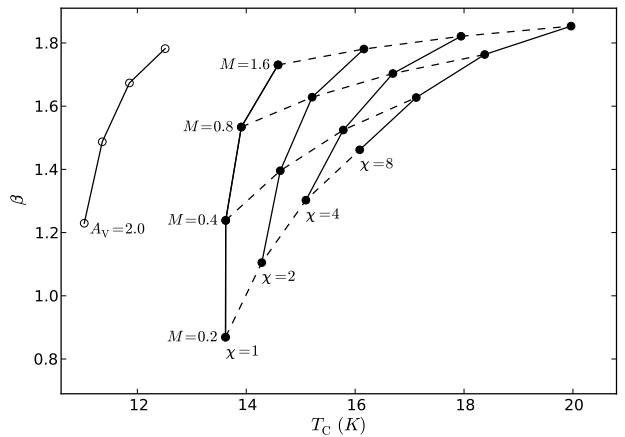}
\caption{
($T_{\rm C}$, $\beta_{\rm Obs}$) values in synthetic observations of
Bonnor-Ebert spheres illuminated by external radiation fields of
different intensity. The cloud masses are 0.2, 0.4, 0.8, and
1.6\,$M_{\sun}$ and the ISRF is scaled by factors $\chi$=1.0, 2.0,
4.0, and 8.0, as indicated in the figure.
The open circles show the case when the ISRF ($\chi$=1.0) is
attenuated by an external layer with $A_{\rm V}=2.0^{\rm m}$ (same as in
Fig.~\ref{fig:final}).
}
\label{fig:ISRF}%
\end{figure}

For large surveys \citep[e.g.][]{Planck2011a}, one further
consideration is the different intensity of the radiation field that
is heating each cloud. Figure~\ref{fig:final} already showed two
cases, one for the normal ISRF, and one for the ISRF extincted by
$A_{\rm V}=2^{\rm m}$. To extend this test to higher radiation field
intensities, Fig.~\ref{fig:ISRF} shows the ($T_{\rm C}$,$\beta_{\rm
Obs}$) values for the Bonnor-Ebert models when the radiation field
intensity is multiplied by factors $\chi$=1.0, 2.0, 4.0, and 8.0. The
correlation between $T_{\rm C}$ and $\beta_{\rm Obs}$ is positive
whether one is comparing clouds of different mass or clouds of given
mass but subjected to different levels of isotropic external radiation
field.
If the external heating is not isotropic, the situation corresponds
qualitatively to two layer models where the temperature difference
between the layers, $\Delta T$, is changing from source to source.
This could again lead in observations to a negative correlation
between the temperature and the spectral index.

The result of the above discussion is that if the real $\beta(T)$
relation is flat, the observed $\beta_{\rm Obs}(T_{\rm C})$ relations
will be very different depending on the set of sources examined. For
clouds heated mainly by an external isotropic radiation field, the
observed $\beta_{\rm Obs}$ should be an increasing function of the
colour temperature, i.e., the correlation between $T_{\rm C}$ and
$\beta_{\rm Obs}$ is positive. When internal or anisotropic external
heating dominates, as in very active star forming regions, $\beta_{\rm
Obs}(T_{\rm C})$ should be a decreasing function of temperature, i.e.,
the correlation is negative. These are merely the qualitative
conclusions and detailed modelling is needed to estimate how the
$\beta(T)$ relation may be modified by the temperature mixing.
Only when models are constructed for individual sources does it
become possible to take into account all the relevant factors like
the precise source geometry, the anisotropies of the local radiation
field, and the location and luminosity of the relevant point sources.

\section{Conclusions}  \label{sect:conclusions}

Even when the grain optical properties are independent of the
temperature, the line-of-sight temperature variations can cause a
correlation between the colour temperature $T_{\rm C}$ and the
observed spectral index $\beta_{\rm Obs}$ that can be either positive
or negative.  The sign of the correlation depends on the nature of the
clouds whose physical differences lead to the dispersion of the
observed $T_{\rm C}$ and $\beta_{\rm Obs}$ values. Examination of a
set of spherical model clouds confirmed the result of
\cite{Malinen2011}, a positive correlation between $T_{\rm C}$ and
$\beta_{\rm Obs}$ in the case of externally heated clouds.
The correlation remains positive when the clouds are heated by
isotropic radiation fields of different intensity. However, internal
heating sources and anisotropic external heating can make the
correlation negative. Quantitative estimates cannot be derived without
detailed modelling and the knowledge of the exact nature of the
sources observed.

\begin{acknowledgements}
MJ and NY acknowledge the support of the Academy of Finland Grant No.
127015 and 250741.
\end{acknowledgements}

\bibliography{biblio_v2.0}

\appendix

\section{Analysis of clumps in a MHD model cloud} \label{sect:collins}

\citet{Malinen2011} presented the analysis of synthetic observations
of a model cloud obtained from a 3D MHD run with self-gravity. The
model cloud had a linear size of 10\,pc and the resolution of the maps
was $\sim$0.005\,pc. Their Fig. 15 shows the ($T_{\rm C}, \beta_{\rm
Obs}$) relations for all pixels in the synthetic maps.  The relation
is shown both when the model is heated only by an external radiation
that corresponds to the local interstellar radiation field
\citep{Mathis1983} and when internal heating sources were added to
gravitationally bound cores.  The masses of the cores are in the range
of $\sim$1--39$M_{\sun}$. The luminosities of the added internal
sources were determined by the core mass and are in the range of
$\sim$0.3-120 solar luminosities. For the details of those
calculations, see \citet{Malinen2011}. The colour temperatures and
spectral indices were based on modified black body fits to data at
100\,$\mu$m, 160\,$\mu$m, 250\,$\mu$m, 350\,$\mu$m, and 500\,$\mu$m,
with the data convolved with a gaussian with FWHM equal to 20 pixels. 

We repeat the analysis using the same synthetic surface brightness
observations but concentrating on the locations of the gravitationally
bound cores. The model contains 40 such cores but internal radiation
sources were added only in 34 cores that were sufficiently resolved in
the automatic mesh refinement (AMR) calculations. Therefore, we
concentrate on the ($T_{\rm C}, \beta_{\rm Obs}$) values towards those
34 cores, before and after the addition of the internal heating.

\begin{figure}
\centering
\includegraphics[width=8.6cm]{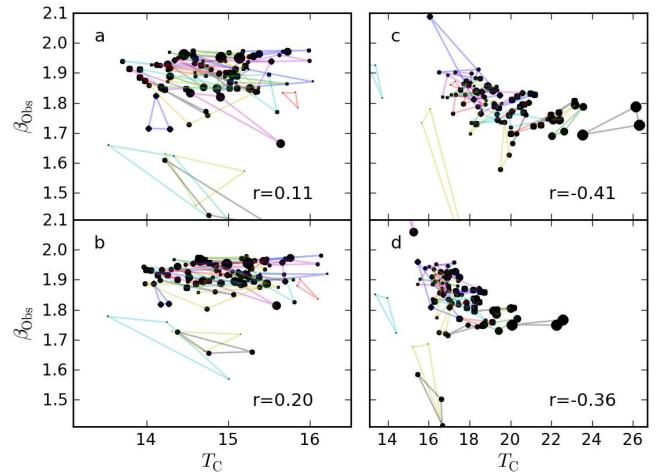}
\caption{
The observed spectral indices and colour temperatures towards the
self-gravitating cores in a model presented in \citet{Malinen2011}.
The frames $a$ and $b$ show the situation without and the frames $c$
and $d$ with heating sources inside the cores. The assumed cloud
distance is 500\,pc in frames $a$ and $c$ and 1\,kpc in the frames $b$
and $d$. The symbol sizes correspond to the mass of the core
($\sim$1-39\,$M_{\sun}$). The surface brightness maps were calculated
for three orthogonal directions. For each core, the lines (of random
colour) connect the values obtained for three different directions of
observation. The linear correlation coefficient between $T_{\rm
C}$ and $\beta_{\rm Obs}$ (including all the plotted points) is given
in the lower right hand corner of each frame.
}
\label{fig:collins_TB}%
\end{figure}

Unlike \cite{Malinen2011}, we examine the data for three orthogonal
viewing direction. Also, we only use wavelengths from 160\,$\mu$m to
500\,$\mu$m. The 100\,$\mu$m is affected by the small grain emission
and is often omitted for the analysis of the large grains. We assume
a cloud distance of 500\,pc or 1.0\,kpc and analyse the surface
brightness values obtained for a single beam towards each of the
cores. The positions are known from the analysis of the 3D density
cube \cite[see][]{Malinen2011}. The beam size is 40$\arcsec$ that,
depending on the distance, corresponds to 20 or 40 pixels. 

The derived spectral index and colour temperature values are shown in
Fig.~\ref{fig:collins_TB}. These are in qualitative agreement with the
results we obtained for spherical models. For externally heated cores
(frames $a$ and $b$) the correlation between $T_{\rm C}$ and
$\beta_{\rm Obs}$ is positive. The correlation is very weak but
consistent with the \cite{Malinen2011} plots that contained data for
the whole extent of the model cloud. The correlation coefficients are
shown in the figure.
With internal heating sources (frames $c$ and $d$), the correlation
between $T_{\rm C}$ and $\beta_{\rm Obs}$ is clearly negative.
Furthermore, the relation appears to be non-linear and reminiscent of
the error bananas produced by noise \cite{Shetty2009b}. However, no
noise was added to these surface brightness values and the scatter in
Fig.~\ref{fig:collins_TB} is caused entirely by the real temperature
variations. 
The ($T_{\rm C}$, $\beta_{\rm Obs}$) values do not show very
significant differences between the viewing directions (i.e., the
triangles in Fig.~\ref{fig:collins_TB} are relatively small). This is
natural when the emission is optically thin and most of the emission
originates in a single core along the line of sight. If some lines of
sight had crossed cores of different temperature, more significant
variation would have been observed. Similarly, at least for this
particular model, the results are not sensitive to the resolution of
the observations.

\end{document}